\documentstyle[12pt,epsfig]{article}
\newcommand{\cl}{\centerline}

\def\be{\begin{equation}}
\def\ee{\end{equation}}
\def\bea{\begin{eqnarray}}
\def\eea{\end{eqnarray}}

\begin{document}

\begin{titlepage}
\setlength{\textwidth}{5.0in}
\setlength{\textheight}{7.5in}
\setlength{\parskip}{0.0in}
\setlength{\baselineskip}{18.2pt}
\hfill {\tt SOGANG-HEP 270/00}
\begin{center}
\cl{\Large{{\bf Global embeddings of scalar-tensor theories}}}\par 
\cl{\Large{{\bf in (2+1)-dimensions}}}\par
\end{center}
\vskip 0.5cm
\begin{center}
{Soon-Tae Hong$^{a}$, Won Tae Kim$^{b}$, Yong-Wan Kim$^{c}$, 
and Young-Jai Park $^{d}$}\par
\end{center}
\vskip 0.4cm
\begin{center}
{Department of Physics and Basic Science Research Institute,}\par
{Sogang University, C.P.O. Box 1142, Seoul 100-611, Korea}\par
\end{center}
\vskip 0.5cm
\cl{\today}
\vfill
\begin{center}
{\bf ABSTRACT}
\end{center}
\begin{quotation}

We obtain (3+3)- or (3+2)-dimensional global flat embeddings of four 
uncharged and charged scalar-tensor theories with the parameters $B$
or $L$ in the (2+1)-dimensions, which are the non-trivially modified 
versions of the Ba\~nados-Teitelboim-Zanelli (BTZ) black holes.  
The limiting cases $B=0$ or $L=0$ exactly are reduced to the Global 
Embedding Minkowski Space (GEMS) solution of the BTZ black holes.


\par
\noindent
--------------------------------------\\
\noindent
$^a$sthong@ccs.sogang.ac.kr\\
\noindent
$^b$wtkim@ccs.sogang.ac.kr\\
\noindent
$^c$ywkim65@netian.com\\
\noindent
$^d$yjpark@ccs.sogang.ac.kr\\
\end{quotation}
\end{titlepage}

\section{Introduction}
After Unruh's work \cite{unr}, it has been known that a thermal
Hawking effect on a curved manifold \cite{hawk75} can be looked
at as an Unruh effect in a higher flat dimensional space-time.
According to the GEMS approach \cite{kasner}, several authors 
\cite{des,bec} recently have shown that this approach could yield a unified 
derivation of temperature for various curved manifolds such as the rotating 
BTZ \cite{btz1,cal,mann93}, the Schwarzschild \cite{sch} 
together with its anti-de Sitter (AdS) extension, the Reissner-Nordstr\"om 
(RN) \cite{rn,pel}, and the RN-AdS \cite{kps99}. 

On the other hand, since the pioneering work in 1992, 
the (2+1)-dimensional BTZ black hole \cite{btz1} has become a useful 
model for realistic black hole physics \cite{cal}.  Moreover, significant 
interest in this model have recently increased with the novel discovery that 
the thermodynamics of higher dimensional black holes can often be 
interpreted in terms of the BTZ solution \cite{high}.
It is therefore interesting to study the geometry of (2+1)-dimensional 
black holes and their thermodynamics through further investigation.
Very recently we have analyse the Hawking and Unruh effects of the
(2+1)-dimensional black holes in terms of the GEMS approach \cite{hkp}.
As a result, we have obtained the novel global higher 
dimensional flat embeddings of the (2+1)-dimensional static, rotating, 
and charged de Sitter (dS) black holes, which are the counterpart of
the usual BTZ black holes as well as the charged static BTZ one.

In this paper we will futher analyse the (2+1)-dimensional scalar-tensor 
(ST) theories \cite{lemos} as an alternative theory of gravity in three 
space-time dimensions in terms of the GEMS approach. 
As you may know three dimensional 
vacuum general relativity (GR) admits no black hole but rather a trivial 
locally flat (globally conical) solution. One has to either couple matter 
to GR, or consider alternative vacuum (or non-vacuum) gravitational theories 
in order to get black hole solutions.  Motivated by this, we will consider 
the GEMS of the new black hole solutions in GR coupled to the vacuum ST 
theories \cite{chaman}, which are modifications of the BTZ black hole by an 
asymptotically constant scalar.

In section 2, we will consider the novel GEMS of the two uncharged 
(2+1)-dimensional ST theories, which have the usual BTZ black hole 
as a substructure.  In section 3, we will also generalize these ST 
theories to the charged cases.

\section{GEMS of uncharged scalar-tensor theories}

In three dimensions, the ST black holes have been obtained in 
Ref. \cite{lemos}. The most general action coupled to 
a scalar can be written as \cite{luis}
\begin{equation}
S=\int  d^3x {\sqrt{-g}} [C(\phi)R-{\omega(\phi)}(\nabla\phi)^2 +V(\phi)], \label{act}
\end{equation}
where $R$ is the scalar curvature,  and $V(\phi)$ is a
potential function for $\phi$. $C(\phi)$ and $\omega(\phi)$ are
collectively known as the coupling functions. 

On the other hand, the field equations for the action (\ref{act}) with 
$C(\phi)=\phi$, which is a choice for the ST theories without loss of 
generality, can be obtained by varying (\ref{act}) with respect to
the metric and scalar fields, respectively, as follows
\be
\phi R_{\mu\nu} = \omega \nabla_{\mu}\phi\nabla_{\nu}\phi - g_{\mu\nu} V +g_{\mu\nu}\nabla^{2}\phi
+\nabla_{\mu}\nabla_{\nu}\phi, 
\label{eoma}
\ee
\be
2\omega\nabla^2\phi + \frac{dV}{d\phi} +\frac{d\omega}{d\phi}
(\nabla \phi )^{2} +R =  0. \label{eomb} 
\ee

The special cases to (\ref{act}) in three dimensions were previously
considered by a number of authors. The first example is the static 
BTZ black hole solution of $C(\phi)=1$, $\omega(\phi)=0$, and
$V(\phi)=2{\Lambda}$ \cite{btz1}. The second example corresponds
to the same $C(\phi)$ as above, but with a non-trivial $\phi$,
$\omega(\phi)=4$ and $V(\phi)=2{\Lambda}e^{b\phi}$, for which the
static  black hole solutions have been previously derived in 
Ref. \cite{chaman}. These examples have the condition $C(\phi)=1$, for
which the metric coupling to matter is the Einstein metric.  In the 
ST theories, this is no longer true for the non-trivial case of 
$C(\phi)\neq 1$, and the gravitational force is governed by a mixture 
of the metric and scalar fields.

We now look for the GEMS of the ST gravity theories described
by field equations (\ref{eoma}) and (\ref{eomb}), which have been already 
analysed by Chan \cite{chan}.  

\subsection{Case I: $\phi = r/(r-3B/2)$}

Let us consider the action and the choice of a scalar field  
\bea
{\cal L} &=& \phi R-{2\over 1-\phi}(\nabla\phi)^2 + 2(3 -
3\phi + \phi^2)\Lambda\phi + {8M\over 27B^2}(1-\phi)^3,
\nonumber\\
\phi &=& {r\over r-{3B\over 2}}, 
\label{lagphi1}
\eea
whose solution is given as
\bea
ds^2 &=& N^{2}dt^2 - N^{-2}{dr^2} - r^2d\theta^2,
\label{esolc} \\
N(r) &=& -M+\frac{MB}{r}+\frac{r^2}{l^{2}}, \label{esolc1}
\eea
where $l^{2}=\Lambda^{-1}$ and $M$ is the positive mass parameter 
calculated using the quasilocal mass \cite{btz1, pel,broyor}.  
Here one notes that the metric looks like the Schwarzschild-AdS metric.  
If $\Lambda=0$, the metric is exactly the same form as the four dimensional 
Schwarzschild case. 

To study the metric (\ref{esolc}) it is convenient to define 
the radial coordinate $r$ as $r\equiv 1/x$. 
Then, the lapse function (\ref{esolc1}) can be rewritten as
\bea
N &=& \frac{M}{x^2}\left(\frac{1}{M l^2}-y_B(x)\right), \nonumber\\
y_B(x) &=& -Bx^3+x^2.
\eea
Note that the parameter $B$ may have either positive or negative values.
The positions of event horizons obtained from $N=0$ can be now read 
off in Fig. 1 from cross sectional curve formed by the surface $y_B(x)$ 
\footnote{In Fig.1 the parameter $B$ is regarded as a continuous 
variable and the limit of $x\rightarrow 0$ corresponds to $r\rightarrow\infty$.  
By choosing a plane with constant $B$, one can easily see that a curve is defined on 
the $(x,y_B)$-plane.  Note that for a fixed negative $B$ there exists only one 
intersection of $x$ associated with the value $\frac{1}{Ml^2}$.} 
and a $(x,B)$-plane associated with a given value $\frac{1}{Ml^2}$.
Moreover, the slope of the curve $y_B(x)$ at intersections along 
the abscissa on a constant $B$ plane gives the surface gravity of the horizon,
which is $k_{H}\equiv\frac{1}{2}\frac{dN}{dr}\mid_{N=0}=M\frac{dy_B}{dx}$.
The positive $B$ region of the graph contains a curve of maximum value, $\frac{4}{27B^2}$, 
along the ordinate $B$.  Thus, when satisfied with $\frac{1}{Ml^2} \leq \frac{4}{27B^2}$, 
there exist two intersections, the outer and inner event horizons, $r_+$ and 
$r_-$, respectively. An extremal black hole appears at the point 
$x=\frac{2}{3B}$ coinciding with $r_+$ and $r_-$ \cite{chaman}. 
On the other hand, for negative $B$ there is only one event horizon 
for any choice of $\frac{1}{Ml^2}$. 

Now, let us consider the GEMS approach to embed this curved spacetime into a
higher dimensional flat one. We restrict ourselves to the region 
of $r>r_+$ according to the usual GEMS embedding \cite{kasner,des,hkp}.
\begin{figure}
\begin{center}
\leavevmode
\epsfysize=5.5cm 
\epsfbox{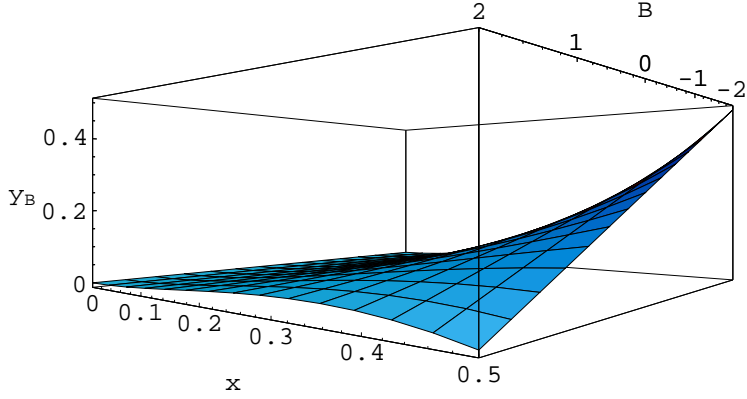}\\ 
\end{center}
\caption[fig1]
{Graph of $y_B(x)=-Bx^3+x^2$: For a given constant $\frac{1}{Ml^2}$, it shows 
that there exist event horizons along the abscissa $x$ on a constant $B$ plane.}
\label{fig1}
\end{figure}

First, for the case of positive $B$ the GEMS embedding is obtained by comparing 
the 3-metric in Eq. (\ref{esolc1}) with $ds^{2}=\eta_{ab}dz^{a}dz^{b}$, where 
$(a,b=0,\cdots,5)$ and $\eta_{ab}=diag(+,-,-,-,+,+)$.  Now, let us find the 
$-r^2d\theta^2$ term in the 3-metric by introducing two coordinates 
$(z^{3}, z^{4})$ in Eq. (\ref{mbtzct}) (see below), 
giving $-(dz^3)^2+(dz^4)^2=-r^2d\theta^2+\frac{l^2}{r_{+}^2}
dr^2$.  Then, in order to obtain the $N^2dt^2$ term, we make an ansatz of two
coordinates, $(z^{0}, z^{1})$ in Eq. (\ref{mbtzct}), which, together with
the above $(z^3, z^4)$, yields 
\begin{eqnarray} 
\label{temp}
&&(dz^0)^2-(dz^1)^2-(dz^3)^2+(dz^4)^2 \nonumber \\
&&=N^2dt^2-\left(k_{H}^{-2}\frac{(-\frac{MB}{2r^{2}}+\frac{r}{l^{2}})^{2}}
{(-M+\frac{MB}{r}+\frac{r^{2}}{l^{2}})}-\frac{l^{2}}{r_{+}^{2}}\right)dr^2
-r^2d\theta^2,
\end{eqnarray}
where the Hawking-Bekenstein horizon surface gravity is given by
\be
k_{H}=\frac{r_{+}}{l^{2}}-\frac{Br_{+}}{2l^{2}(r_{+}-B)}.
\ee
Since the combination of $N^{-2}dr^{2}$ and $dr^2$ terms in Eq. (\ref{temp}) 
can be separated into a positive definite part and a negative one as follows 
\begin{eqnarray}
\label{temp2}
&& \left(k_{H}^{-1}\frac{lN_{1}^{1/2}(B)}{2r_{+}^{2}r^{3/2}(r_{+}-B)
[r_{+}r(r+r_{+})-B(r^{2}+r_{+}r+r_{+}^{2})]^{1/2}}\right)^{2}
\nonumber\\
& &-\left(k_{H}^{-1}\frac{lN_{2}^{1/2}(B)}{2r_{+}^{2}r^{3/2}(r_{+}-B)
[r_{+}r(r+r_{+})-B(r^{2}+r_{+}r+r_{+}^{2})]^{1/2}}\right)^{2}
\nonumber \\
&& \equiv (dz^2)^2 - (dz^5)^2,
\end{eqnarray}
where 
\bea
N_{1}(B)&=&\frac{B^{2}r_{+}^{5}}{l^{4}}[r_{+}^{3}(r^{2}+r_{+}r+r_{+}^{2})
+9r^{4}(r+r_{+})+21r_{+}^{2}r^{3}],
\nonumber\\
N_{2}(B)&=&\frac{Br_{+}^{4}}{l^{4}}[(8r_{+}^{2}+14B^{2})r_{+}^{2}r^{3}
+B^{2}r_{+}^{3}(r^{2}+r_{+}r+r_{+}^{2})\nonumber\\
& &+(4r_{+}^{2}+5B^{2})r^{4}(r+r_{+})],
\label{nn12}
\eea
we can obtain the flat global embeddings of the corresponding 
curved 3-metric as
\begin{eqnarray}
ds^{2}&=&(dz^{0})^2-(dz^{1})^2
  -(dz^{2})^2-(dz^{3})^{2} +(dz^{4})^{2}+(dz^{5})^{2} \nonumber \\
      &=& N^2dt^2-N^{-2}dr^2-r^2d\theta^2.
\end{eqnarray}

As a result, the desired coordinate transformations to the (3+3)-dimensional 
AdS GEMS are obtained for $r\geq r_{+}$ as
\bea 
z^{0}&=&k_{H}^{-1}(-M+\frac{MB}{r}+\frac{r^{2}}{l^{2}})^{1/2}\sinh k_{H}t,\nonumber \\
z^{1}&=&k_{H}^{-1}(-M+\frac{MB}{r}+\frac{r^{2}}{l^{2}})^{1/2}\cosh k_{H}t,\nonumber \\
z^{2}&=&k_{H}^{-1}\int {\rm d}r\frac{lN_{1}^{1/2}(B)}{2r_{+}^{2}r^{3/2}(r_{+}-B)
[r_{+}r(r+r_{+})-B(r^{2}+r_{+}r+r_{+}^{2})]^{1/2}},
 \nonumber \\
z^{3}&=&\frac{l}{r_{+}}r\sinh \frac{r_{+}}{l}\theta,\nonumber\\
z^{4}&=&\frac{l}{r_{+}}r\cosh \frac{r_{+}}{l}\theta,\nonumber\\
z^{5}&=&k_{H}^{-1}\int {\rm d}r\frac{lN_{2}^{1/2}(B)}{2r_{+}^{2}r^{3/2}(r_{+}-B)
[r_{+}r(r+r_{+})-B(r^{2}+r_{+}r+r_{+}^{2})]^{1/2}}.
\label{mbtzct}
\eea

In static detectors ($\theta$, $r=$ const) described by a fixed
point in the ($z^{2}$, $z^{3}$, $z^{4}$, $z^{5}$) hyper-plane, 
one can have constant 3-acceleration
\be
a=\frac{\frac{r}{l^{2}}-\frac{MB}{2r^{2}}}
{(-M+\frac{MB}{r}+\frac{r^{2}}{l^{2}})^{1/2}},
\ee
and constant accelerated motion in ($z^{0}$,$z^{1}$) plane
with the Hawking temperature
\be
2\pi T=a_{6}=\frac{\frac{r_{+}}{l^{2}}-\frac{MB}{2r_{+}^{2}}}{(-M+\frac{MB}{r}+\frac{r^{2}}{l^{2}})^{1/2}}.
\ee
Here one notes that the above Hawking temperature 
is also given by the relation \cite{hawk75, bek73}:
\be
T=\frac{1}{2\pi}\frac{k_{H}}{g_{00}^{1/2}}. 
\label{g00}
\ee 
One can easily check that, in the limit of $B=0$ where the spacelike 
$z^{2}$ and timelike $z^{5}$ coordinates in Eq. (\ref{mbtzct}) vanish, the 
above (3+3)-dimensional coordinate transformations are exactly reduced 
to the (2+2)-dimensional GEMS of the usual BTZ case \cite{des,hkp}.

We now see how the scalar-tensor solution, which is a modified version of the 
BTZ, yields a finite Unruh area due to the periodic identification of $\theta$
mod $2\pi$.  The Rindler horizon condition $(z^1)^2- (z^0)^2 = 0$ implies 
$r=r_{+}$ and the embedding constraints yield $z^{2}=f_{1}(r)$, 
$z^{5}=f_{2}(r)$, and $(z^4)^2 - (z^3)^2 =l^2$ where $f_{1}(r)$ and $f_{2}(r)$ 
can be read from Eq. (\ref{mbtzct}).  The area of the Rindler horizon is 
now described as  
\[
\int {\rm d}z^{2}{\rm d}z^{3}{\rm d}z^{4}{\rm d}z^{5}
\delta(z^{2}-f_{1}(r))\delta(z^{5}-f_{2}(r))
\delta([(z^4)^2-(z^3)^2]^{1/2}-l),
\]
which, after performing trivial integrations over $z^2$ and $z^5$, 
yields the desired entropy of the scalar-tensor theory as
\bea
& &\int^{l\sinh (\pi r_{+} /l)}_{-l \sinh (\pi r_{+} /l)}{\rm d}z^{3}
     \int^{[(z^3)^2+l^2]^{1/2}}_{0}{\rm d}z^{4}
     \delta([(z^4)^2-(z^3)^2]^{1/2}-l)\nonumber\\
&=&\int^{l\sinh (\pi r_{+} /l)}_{-l \sinh (\pi r_{+} /l)}{\rm d}z^{3}
     \frac{l}{[l^2 +(z^3)^2]^{1/2}}=2\pi r_{+}(B),
\eea
which reproduces the entropy $2\pi r_{H}$ of the uncharged BTZ case 
in the limit $B=0$.

Next, for the case of $B<0$, since $N_2(B)$ is an odd function of $B$,
the combination of $N^{-2}dr^2$ and $dr^2$ terms in Eq. (\ref{temp})
can be written by introducing only one extra space
\footnote{By a simple test with $B<0$, we can show that Eq. (\ref{temp3}) 
is really monotonic decreasing function, and thus can be defined as 
a spacelike variable.}
dimension $z^{'2}$
as follows
\begin{eqnarray}
\label{temp3}
&& -\left(k_{H}^{-1}\frac{l(N_{2}-N_{1})^{1/2}(B)}
{2r_{+}^{2}r^{3/2}(r_{+}-B)
[r_{+}r(r+r_{+})-B(r^{2}+r_{+}r+r_{+}^{2})]^{1/2}}\right)^{2} \nonumber\\
&& \equiv -(dz^{'2})^2.
\end{eqnarray}
Then, we can obtain the following flat embedding of the corresponding curved 
3-metric as
\bea
ds^2 &=& (dz^{0})^2-(dz^{1})^2-(dz^{'2})^2-(dz^{3})^{2}+(dz^{4})^{2} \nonumber \\
     &=& N^2dt^2-N^{-2}dr^2-r^2d\theta^2.
\eea
As a result, the desired coordinate transformations to the (3+2)-dimensional 
GEMS are for $r>r_+$
\be
z^{'2}=k_H^{-1}\int {\rm d}r \frac{l(N_2-N_1)^{1/2}(B)}
{2r_{+}^{2}r^{3/2}(r_{+}-B)
[r_{+}r(r+r_{+})-B(r^{2}+r_{+}r+r_{+}^{2})]^{1/2}},
\ee
while $(z^0, z^1, z^3, z^4)$ are of those forms in Eq. (\ref{mbtzct}).
Similar to the previous $B>0$ case, one can easily obtain the desired entropy of the
ST theory as $2\pi r(B)$, where $r$ is only one event horizon in this case.

It seems appropriate to comment on the minimal extra dimensions 
needed for a desired GEMS.  As you may know, spaces of constant curvature can 
be embedded into flat space with only single extra dimension.  This 
is seen in our previous work \cite{hkp} for the static and rotating BTZ 
cases, which are embedded in the (2+2)-dimensional spaces.  On the other hand, 
since the scalar-tensor solution is Schwarzschild-like \cite{des,kps99}, 
we have introduced (1+2) or (1+1) extra dimensions for the desired GEMS 
with the positive or negative $B$, respectively.  
In the next section, we will also obtain similar results for the charged 
scalar-tensor theories. 

\subsection{Case II: $\phi=r^2/(r^2-2L)$}

Next, an another choice of an asymptotically constant scalar yields
\bea
{\cal L} &=& \phi R-{4\phi-1\over 2\phi(1-\phi)}(\nabla\phi)^2
+ {M\over 2L} + 6\left(2\Lambda-{M\over 2 L}\right)\phi \nonumber\\
& &+18\left(-\Lambda+{M\over 4L}\right)\phi^2 + 2\left(4\Lambda -
{M\over L}\right)\phi^3, \nonumber\\ 
\phi &=&{r^2\over r^2-2L}, \label{esole1}
\eea
to yield the solution
\bea
ds^2 &=& N^{2}dt^2 - N^{-2}{dr^2} - r^2d\theta^2, 
\nonumber\\
N(r) &=& -M+\frac{ML}{r^{2}}+\frac{r^{2}}{l^{2}},
\label{metric2}
\eea
where $l^{2}=\Lambda^{-1}$.  The metric has a curvature singularity
at $r=0$, and the scalar and its potential both diverge at $r^2={2L}$
with $L>0$.  Note that the only case of $L>0$ is physically meaningful
since we require to have the positive $r$.

Now, defining the radial coordinate $r$ as $r=1/x$ 
as in the previous subsection, the lapse function can be rewritten as
\bea
N &=& \frac{M}{x^2}\left( \frac{1}{Ml^2} - y_L(x)\right),\\
y_L(x) &=& -Lx^4+x^2.
\eea
As shown in Fig. 2, for a specific value of $\frac{1}{Ml^2}$, there exist 
two event horizons, $r_+$ and $r_-$ on a $(x, y_L)$-plane with a constant $L$, 
if satisfied $\frac{1}{Ml^2}\leq\frac{1}{4L}$. Here, 
the maximum value of $\frac{1}{4L}$ is obtained at $x=\frac{1}{\sqrt{2L}}$ 
($r=\sqrt{2L}$).
\begin{figure}
\begin{center}
\leavevmode\epsfysize=5.5cm \epsfbox{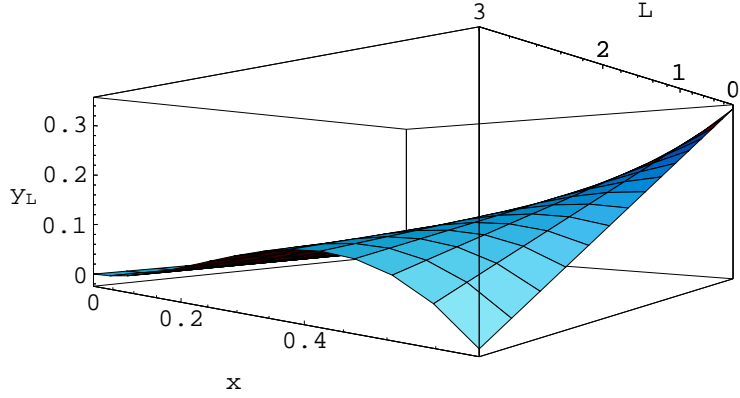}\\ 
\end{center}
\caption[fig2]
{Graph of $y_L(x)=-Lx^4+x^2$: For a given constant $\frac{1}{Ml^2}$, it 
shows that there exist event horizons along the abscissa $x$ on a 
constant $L$ plane.}
\label{fig2}
\end{figure}
Moreover, the extremal limit is $4L=Ml^2$ at $x=1/\sqrt{2L}$ ($r=\sqrt{2L}$) 
coinciding with $r_+$ and $r_-$.  In this limit, the third term in the Lagrangian 
becomes $2\Lambda$, while fourth, fifth and sixth terms all vanish.  
Here note that, in the limit $L=\frac{J^{2}}{4M}$, 
the solution seems to be related to a rotationg BTZ black hole.  
However, since the 3-metric (\ref{metric2}) does not contain a shift 
function, our ST theory does not allow such a rotating BTZ solution.  

After similar algebraic manipulation for the region of $r>r_+$ 
by following the previous steps described in Sec. 2.1, 
we obtain the desired coordinate transformations to the (3+3)-dimensional 
AdS GEMS, $ds^{2}=(dz^{0})^2-(dz^{1})^2 -(dz^{2})^2 -(dz^{3})^2+
(dz^{4})^{2}+(dz^{5})^{2}$, which are obtained for $r\geq r_{+}$:
\bea 
z^{0}&=&k_{H}^{-1}(-M+\frac{ML}{r^{2}}+\frac{r^{2}}{l^{2}})^{1/2}\sinh k_{H}t,\nonumber \\
z^{1}&=&k_{H}^{-1}(-M+\frac{ML}{r^{2}}+\frac{r^{2}}{l^{2}})^{1/2}\cosh k_{H}t,\nonumber \\
z^{2}&=&k_{H}^{-1}\int {\rm d}r\frac{lN_{3}^{1/2}(L)}{r_{+}^{2}r^{2}
(r_{+}^{2}-L)[r_{+}^{2}r^{2}-L(r^{2}+r_{+}^{2})]^{1/2}},
\nonumber \\
z^{3}&=&\frac{l}{r_{+}}r\sinh \frac{r_{+}}{l}\theta,\nonumber\\
z^{4}&=&\frac{l}{r_{+}}r\cosh \frac{r_{+}}{l}\theta,\nonumber\\
z^{5}&=&k_{H}^{-1}\int {\rm d}r\frac{lN_{4}^{1/2}(L)}
{r_{+}^{2}r^{2}(r_{+}^{2}-L)[r_{+}^{2}r^{2}-L(r^{2}+r_{+}^{2})]^{1/2}},
\label{trfmucl}
\eea
where the Hawking-Bekenstein horizon surface gravity is given by 
\be
k_{H}=\frac{r_{+}}{l^{2}}-\frac{L r_{+}}{l^{2}(r_{+}^{2}-L)},
\ee
and
\bea
N_{3}(L)&=&\frac{L^{2}r_{+}^{6}}{l^{4}}
[r_{+}^{2}(r^{4}+r_{+}^{2}r^{2}+r_{+}^{4})+5r^{4}(r^{2}+r_{+}^{2})
+3r_{+}^{2}r^{4}],\nonumber\\
N_{4}(L)&=&\frac{Lr_{+}^{4}}{l^{4}}
[(r_{+}^{4}+3L^{2})r_{+}^{2}r^{4}+L^{2}r_{+}^{2}
(r^{4}+r_{+}^{2}r^{2}+r_{+}^{4})
\nonumber\\
& &+(2r_{+}^{4}+3L^{2})r^{4}(r^{2}+r_{+}^{2})].
\eea

In static detectors ($\theta$, $r=$ const) described by a fixed
point in the ($z^{2}$, $z^{3}$, $z^{4}$, $z^{5}$) hyper-plane, 
one can have constant 3-acceleration
\be
a=\frac{\frac{r}{l^{2}}-\frac{ML}{r^{3}}}
{(-M+\frac{ML}{r^{2}}+\frac{r^{2}}{l^{2}})^{1/2}},
\ee
and constant accelerated motion in ($z^{0}$,$z^{1}$) plane 
with the Hawking temperature
\be
2\pi T=a_{6}=\frac{\frac{r_{+}}{l^{2}}-\frac{ML}{r_{+}^{3}}}
{(-M+\frac{ML}{r^{2}}+\frac{r^{2}}{l^{2}})^{1/2}}.
\ee
Here one notes that the above Hawking temperature 
is also given by the relation (\ref{g00}).

Similar to the previous case, in the limit of $L=0$, where the spacelike 
$z^{2}$ and timelike $z^{5}$ coordinates in Eq. (\ref{trfmucl}) vanish, 
the (3+3)-dimensional coordinate transformations are exactly reduced 
to the (2+2)-dimensional GEMS of the usual BTZ case \cite{des,hkp}.  
We also obtain the entropy $2\pi r_{+}(L)$ of the 
scalar-tensor theory with $\phi = r^{2}/(r^{2}-2L)$, which reproduces the 
uncharged static BTZ entropy $2\pi r_{H}$ \cite{des,hkp} in the $L=0$ limit.

\section{GEMS of charged scalar-tensor theories}

\subsection{Case I: $\phi = r/(r-3B/2)$}

Now consider the charged scalar-tensor theory for the modified BTZ black hole 
where the 3-metric (\ref{esolc}) is described by the charged lapse
\begin{equation}
N(r) = -M+\frac{MB}{r}+\frac{r^2}{l^{2}}-2Q^{2}\ln r.
\label{ncb}
\end{equation}
Here we only consider the case in which the parameter $B$ is positive
because the analysis for the case of $B<0$ is highly non-trivial due to
the addition of the charged term in contrast to the uncharged cases.

The coordinate transformations to the (3+3)-dimensional AdS GEMS 
$ds^{2}=(dz^{0})^2-(dz^{1})^2 -(dz^{2})^2 -(dz^{3})^2+(dz^{4})^{2}
+(dz^{5})^{2}$ are obtained for $r\geq r_{+}$:
\bea 
z^{0}&=&k_{H}^{-1}(-M+\frac{MB}{r}+
       \frac{r^{2}}{l^{2}}-2Q^{2}\ln r )^{1/2}\sinh k_{H}t,\nonumber \\
z^{1}&=&k_{H}^{-1}(-M+\frac{MB}{r}
       +\frac{r^{2}}{l^{2}}-2Q^{2}\ln r )^{1/2}\cosh k_{H}t,\nonumber \\
z^{2}&=&k_{H}^{-1}\int {\rm  d}r\frac{lN_{1}^{1/2}(B,Q)}
        {2r_{+}^{3/2}r^{3/2}(r_{+}-B)D_{1}^{1/2}(B,Q)},\nonumber\\
z^{3}&=&\frac{l}{r_{+}}r\sinh \frac{r_{+}}{l}\theta,\nonumber\\
z^{4}&=&\frac{l}{r_{+}}r\cosh \frac{r_{+}}{l}\theta,\nonumber\\
z^{5}&=&k_{H}^{-1}\int {\rm  d}r\frac{lN_{2}^{1/2}(B,Q)}
       {2r_{+}^{3/2}r^{3/2}(r_{+}-B)D_{1}^{1/2}(B,Q)},
\label{trfmbc}
\eea
where the Hawking-Bekenstein horizon surface gravity is given by 
\be
k_{H}=\frac{r_{+}}{l^{2}}-\frac{B r_{+}^{2}-2BQ^{2}l^{2}\ln r_{+}}
      {2l^{2}r_{+}(r_{+}-B)}-\frac{Q^{2}}{r_{+}},
\ee
and
\bea
N_{1}(B,Q)&=&4Q^{4}r_{+}^{3}r^{2}(r+r_{+})[r_{+}^{2}+r^{2}(2f+1)]\nonumber\\
&& +\frac{B^{2}r_{+}^{5}}{l^{4}}[r_{+}^{3}
(r^{2}+r_{+}r+r_{+}^{2})+9r^{4}(r+r_{+})+21r_{+}^{2}r^{3}]
\nonumber\\
&& +c_{12}BQ^{2}+c_{16}BQ^{6}+c_{24}B^{2}Q^{4}
+c_{32}B^{3}Q^{2}+c_{36}B^{3}Q^{6},\nonumber\\
N_{2}(B,Q)&=&\frac{4Q^{2}r_{+}^{3}r^{4}}{l^{2}}
(r+r_{+})(2r_{+}^{2}+\frac{r_{+}^{4}+Q^{4}l^{4}}{r_{+}^{2}}f)
\nonumber\\
& &+\frac{Br_{+}^{4}}{l^{4}}[(8r_{+}^{2}+14B^{2})r_{+}^{2}r^{3}
+B^{2}r_{+}^{3}(r^{2}+r_{+}r+r_{+}^{2})\nonumber\\
& &+(4r_{+}^{2}+5B^{2})r^{4}(r+r_{+})]\nonumber\\
& &+c_{14}BQ^{4}+c_{22}B^{2}Q^{2}+c_{26}B^{2}Q^{6}+c_{34}B^{3}Q^{4},
\nonumber\\
D_{1}(B,Q)&=&r_{+}^{2}r(r+r_{+})-Br_{+}(r^{2}+r_{+}r+r_{+}^{2})\nonumber\\
& &-Q^{2}l^{2}(r+r_{+})[rf-B(\ln r_{+}+1)g],
\label{nnd1}
\eea
and the coefficients are given by
\bea
c_{12}&=&\frac{4r_{+}r}{l^{2}}[r_{+}^{3}(2r^{2}+r_{+}^{2}+2r^{2}\ln r_{+})(r^{2}+r_{+}r+r_{+}^{2})\nonumber\\
& &+r_{+}^{3}r^{3}(r+r_{+})(3f+5)+r_{+}^{4}r^{2}(r+r_{+})(\ln r_{+} +1)g],
\nonumber\\
c_{14}&=&4r_{+}^{2}r[r^{2}(r^{2}+r_{+}r+r_{+}^{2})(2\ln r_{+}+3)+2r_{+}^{3}(r+r_{+})\ln r_{+}\nonumber\\
& &+r_{+}^{2}r(r+3r_{+})+r^{3}(r+r_{+})(2\ln r_{+}+5)f\nonumber\\
& &+2r_{+}r^{2}(r+r_{+})(\ln r_{+}+1)g],
\nonumber\\
c_{16}&=&4l^{2}r^{3}(r+r_{+})(\ln r_{+} +1)(2rf+r_{+}g),\nonumber\\
c_{22}&=&\frac{r_{+}^{3}}{l^{2}}[20r^{3}(r^{2}+r_{+}r+r_{+}^{2})(\ln r_{+}+1)
+8r_{+}^{2}r^{3}\ln r_{+}\nonumber\\
& &+4r_{+}^{2}(r^{2}+r_{+}r+r_{+}^{2})(2r+r_{+}\ln r_{+})+3r^{4}(r+r_{+})(3f+4)\nonumber\\
& &+12r_{+}r^{3}(r+r_{+})(\ln r_{+}+1)g],\nonumber\\
c_{24}&=&4r_{+}[r^{3}(r^{2}+r_{+}r+r_{+}^{2})(\ln r_{+} +1)^{2}+3r_{+}^{4}r
\ln r_{+}\nonumber\\
& &+r_{+}^{2}(r^{2}+r_{+}r+r_{+}^{2})(r+r_{+}\ln r_{+})\ln r_{+}\nonumber\\
& &+(r+r_{+})(\ln r_{+}+1)(3r^{4}f+2r^{4}+3r_{+}^{2}r^{2})\nonumber\\
& &+r_{+}r^{3}(r+r_{+})(\ln r_{+}+1)(2\ln r_{+}+5)g],
\nonumber\\
c_{26}&=&4l^{2}r^{3}\frac{r+r_{+}}{r_{+}}(\ln r_{+} +1)^{2}(rf+2r_{+}g),
\nonumber\\
c_{32}&=&\frac{r_{+}^{2}}{l^{2}}
[4r(r_{+}^{2}+3r^{2})(r^{2}+r_{+}r+r_{+}^{2})(\ln r_{+}+1)\nonumber\\
& &+4r_{+}^{2}(2r^{3}+r_{+}^{3})\ln r_{+}
+9r_{+}r^{3}(r+r_{+})(\ln r_{+} +1)g],
\nonumber\\
c_{34}&=&4r^{3}(r^{2}+r_{+}r+r_{+}^{2})
(\ln r_{+}+1)^{2}+4r_{+}^{4}(r+r_{+})(\ln r_{+})^{2}\nonumber\\
& &+8r_{+}^{4}r\ln r_{+}+4r_{+}r^{2}(r+r_{+})(\ln r_{+}+1)^{2}
(r_{+}+3rg),\nonumber\\
c_{36}&=&4l^{2}r^{3}\frac{r+r_{+}}{r_{+}}(\ln r_{+} +1)^{3}g,\nonumber\\
f(r,r_{+})&=&\frac{2r_{+}^{2}\ln (r/r_{+})}{r^{2}-r_{+}^{2}},\nonumber\\
g(r,r_{+})&=&\frac{2r_{+}(r\ln r -r_{+}\ln r_{+})}{(r^{2}-r_{+}^{2})
(\ln r_{+}+1)}.
\eea
Here both $f(r,r_{+})$ and $g(r,r_{+})$ approach to unities as $r$ goes 
to $r_{+}$, due to L'Hospital's rule. 

In static detectors ($\theta$, $r=$ const) described by a fixed
point in the ($z^{2}$, $z^{3}$, $z^{4}$, $z^{5}$) hyper-plane, 
one can have constant 3-acceleration
\be
a=\frac{\frac{r}{l^{2}}-\frac{MB}{2r^{2}}-\frac{Q^{2}}{r}}{(-M+\frac{MB}{r}
+\frac{r^{2}}{l^{2}}-2Q^{2}\ln r)^{1/2}},
\ee
and constant accelerated motion in ($z^{0}$,$z^{1}$) plane with the Hawking 
temperature
\be
2\pi T=a_{6}=\frac{\frac{r_{+}}{l^{2}}-\frac{MB}{2r_{+}^{2}}-\frac{Q^{2}}
{r_{+}}}{(-M+\frac{MB}{r}+\frac{r^{2}}{l^{2}}-2Q^{2}\ln r)^{1/2}}.
\ee
On the other hand, the above Hawking temperature is also given by the 
relation (\ref{g00}).  Note that one can easily check that, since in the 
uncharged limit $Q=0$, $N_{1}(B,Q)$ and $N_{2}(B,Q)$ in Eq. (\ref{nnd1}) are 
exactly reduced to the $N_{1}(B)$ and $N_{2}(B)$ in Eq. (\ref{nn12}), 
respectively, the (3+3)-dimensional coordinate transformations (\ref{trfmbc}) 
are also exactly reduced to the uncharged case (\ref{mbtzct}) having the same 
(3+3)-dimensional GEMS structure in contrast to the usual BTZ case \cite{hkp}.  
Since in this case the metric is Schwarzschild-like, the GEMS structure 
coinsides with that of the (3+1)-dimensional Schwarzschild black hole, 
which needs (1+1) additional extra dimensions to yield the (4+2) GEMS 
structure \cite{des,kps99}.  
Furthermore, in the $B=0$ limit, the transformations (\ref{trfmbc}) are 
exactly reduced to the charged BTZ case \cite{hkp}, which still has 
the (3+3)-dimensional GEMS structure.

\subsection{Case II: $\phi=r^2/(r^2-2L)$}

Now consider the charged scalar-tensor theory with $L>0$ 
for the modified BTZ black hole where the 3-metric (\ref{metric2}) 
is described by the charged lapse function:
\begin{equation}
N(r) = -M+\frac{ML}{r^{2}}+\frac{r^2}{l^{2}}-2Q^{2}\ln r.
\label{cnl}
\end{equation}

The coordinate transformations to the (3+3)-dimensional AdS GEMS, 
$ds^{2}=(dz^{0})^2-(dz^{1})^2 -(dz^{2})^2 -(dz^{3})^2+(dz^{4})^{2}
+(dz^{5})^{2}$ are obtained for $r\geq r_{+}$:
\bea 
z^{0}&=&k_{H}^{-1}(-M+\frac{ML}{r^{2}}+\frac{r^{2}}
      {l^{2}}-2Q^{2}\ln r)^{1/2}\sinh k_{H}t,\nonumber \\
z^{1}&=&k_{H}^{-1}(-M+\frac{ML}{r^{2}}+
      \frac{r^{2}}{l^{2}}-2Q^{2}\ln r)^{1/2}\cosh k_{H}t,\nonumber \\
z^{2}&=&k_{H}^{-1}\int {\rm  d}r\frac{lN_{3}^{1/2}(L,Q)}
        {r_{+}^{2}r^{2}(r_{+}^{2}-L)D_{2}^{1/2}(L,Q)},\nonumber \\
z^{3}&=&\frac{l}{r_{+}}r\sinh \frac{r_{+}}{l}\theta,\nonumber\\
z^{4}&=&\frac{l}{r_{+}}r\cosh \frac{r_{+}}{l}\theta,\nonumber\\
z^{5}&=&k_{H}^{-1}\int {\rm  d}r
    \frac{lN_{4}^{1/2}(L,Q)}{r_{+}^{2}r^{2}(r_{+}^{2}-L)D_{2}^{1/2}(L,Q)},
\eea
where the Hawking-Bekenstein horizon surface gravity is given by 
\be
k_{H}=\frac{r_{+}}{l^{2}}-\frac{L r_{+}^{2}-2LQ^{2}l^{2}
      \ln r_{+}}{l^{2}r_{+}(r_{+}^{2}-L)}-\frac{Q^{2}}{r_{+}},
\ee
and
\bea
N_{3}(L,Q)&=&Q^{4}r_{+}^{6}r^{4}[r_{+}^{2}+r^{2}(2f+1)]\nonumber\\
&&+\frac{L^{2}r_{+}^{6}}{l^{4}}[r_{+}^{2}(r^{4}+r_{+}^{2}r^{2}
  +r_{+}^{4})+5r^{4}(r^{2}+r_{+}^{2})+3r_{+}^{2}r^{4}]
\nonumber\\
&&+d_{12}LQ^{2}+d_{16}LQ^{6}+d_{24}L^{2}Q^{4}
  +d_{32}L^{3}Q^{2}+d_{36}L^{3}Q^{6},\nonumber\\
N_{4}(L,Q)&=&\frac{Q^{2}r_{+}^{6}r^{6}}{l^{2}}(2r_{+}^{2}
+\frac{r_{+}^{4}+Q^{4}l^{4}}{r_{+}^{2}}f)
\nonumber\\
&&+\frac{Lr_{+}^{4}}{l^{4}}[(r_{+}^{4}
+3L^{2})r_{+}^{2}r^{4}+L^{2}r_{+}^{2}(r^{4}+r^{2}r_{+}^{2}+r_{+}^{4})
\nonumber\\
&&+(2r_{+}^{4}+3L^{2})r^{4}(r^{2}+r_{+}^{2})]\nonumber\\
&&+d_{14}LQ^{4}+d_{22}L^{2}Q^{2}+d_{26}L^{2}Q^{6}+d_{34}L^{3}Q^{4},
\nonumber\\
D_{2}(L,Q)&=&r_{+}^{2}r^{2}-L(r^{2}+r_{+}^{2})-Q^{2}l^{2}
[r^{2}f-L(2\ln r_{+}+1)g],
\eea
and the coefficients are given by
\bea
d_{12}&=&\frac{r_{+}^{4}r^{2}}{l^{2}}[2r_{+}^{2}r^{2}(r^{2}+r_{+}^{2})
(2\ln r_{+}+1)+2r_{+}^{2}(r^{4}+r_{+}^{2}r^{2}+r_{+}^{4})\nonumber\\
& &+4r_{+}^{2}r^{4}(f+1)+r_{+}^{4}r^{2}(2\ln r_{+} +1)g],
\nonumber\\
d_{14}&=&r_{+}^{4}r^{2}[4(r^{4}+r_{+}^{2}r^{2}+r_{+}^{4})
        \ln r_{+}+r^{2}(3r^{2}+4r_{+}^{2})\nonumber\\
& &+2r^{4}(2\ln r_{+}+3)f+2r_{+}^{2}r^{2}(2\ln r_{+}+1)g],
\nonumber\\
d_{16}&=&l^{2}r_{+}^{2}r^{4}[2r^{2}f+r_{+}^{2}(2\ln r_{+} +1)g),\nonumber\\
d_{22}&=&\frac{2r_{+}^{4}}{l^{2}}[2(r^{4}+r_{+}^{2}r^{2}+r_{+}^{4})
        (r^{2}+r_{+}^{2}\ln r_{+})
+3r^{4}(r^{2}+r_{+}^{2})(2\ln r_{+}+1)],\nonumber\\
d_{24}&=&r_{+}^{2}[r^{2}(r^{4}+r_{+}^{2}r^{2}+r_{+}^{4})(2\ln r_{+} +1)^{2}
+4r_{+}^{4}(r^{2}+r_{+}^{2}\ln r_{+})\ln r_{+}\nonumber\\
&&+2r^{6}(2\ln r_{+}+1)(2f+1)+2r_{+}^{2}r^{4}(2\ln r_{+}+1)(2\ln r_{+}+3)g],
\nonumber\\
d_{26}&=&l^{2}r^{4}(2\ln r_{+} +1)^{2}(r^{2}f+2r_{+}^{2}g),\nonumber\\
d_{32}&=&\frac{r_{+}^{2}r^{2}}{l^{2}}[(r_{+}^{4}+r_{+}^{2}r^{2}
  +r^{4})+8r^{2}(r^{2}+r_{+}^{2})\ln r_{+}+2r^{2}(2r^{2}+r_{+}^{2})],
\nonumber\\
d_{34}&=&4(r^{4}+r_{+}^{2}r^{2}+r_{+}^{4})(r^{2}+r_{+}^{2}\ln r_{+})
        \ln r_{+}+r_{+}^{2}r^{4}(2\ln r_{+}+1)^{2}\nonumber\\
& &+r^{6}[4(\ln r_{+})^{2}+1]+4r_{+}^{2}r^{4}(2\ln r_{+}+1)^{2}g,
\nonumber\\
d_{36}&=&l^{2}r^{4}(2\ln r_{+}+1)^{3}g.
\eea

In static detectors ($\theta$, $r=$ const) described by a fixed
point in the ($z^{2}$, $z^{3}$, $z^{4}$, $z^{5}$) hyper-plane, 
one can have constant 3-acceleration
\be
a=\frac{\frac{r}{l^{2}}-\frac{ML}{r^{3}}-\frac{Q^{2}}{r}}
{(-M+\frac{ML}{r^{2}}+\frac{r^{2}}{l^{2}}-2Q^{2}\ln r)^{1/2}},
\ee
and constant accelerated motion in ($z^{0}$,$z^{1}$) plane
with the Hawking temperature
\be
2\pi T=a_{6}=\frac{\frac{r_{+}}{l^{2}}-\frac{ML}
{r_{+}^{3}}-\frac{Q^{2}}{r_{+}}}
{(-M+\frac{ML}{r^{2}}+\frac{r^{2}}{l^{2}}-2Q^{2}\ln r)^{1/2}}.
\ee
Similar to the previous case, one can also check that, since in the 
uncharged limit of $Q=0$, $N_{3}(L,Q)$ and $N_{4}(L,Q)$ are exactly reduced 
to the $N_{1}(L)$ and $N_{2}(L)$, respectively, the coordinate transformations 
(\ref{trfmbc}) are also exactly reduced to the uncharged case (\ref{mbtzct}) 
having the same (3+3)-dimensional GEMS structure 
in contrast to the usual BTZ case \cite{hkp}.  
Furthermore, in the $L=0$ limit, the transformations (\ref{trfmbc}) 
are exactly reduced to the charged BTZ case \cite{hkp}, 
which still has the (3+3)-dimensional GEMS structure.

\section{Conclusions}

In conclusion, we have newly analyzed the (2+1)-dimensional 
four uncharged and two charged ST theories with the parameters $B$ or 
$L$ through the GEMS approach, which are the modified versions 
of the usual BTZ black holes.  
First, we have obtained the (3+3)- or (3+2)-dimensional GEMS of the uncharged
ST theories in the (2+1)-dimensions depending on the positive or negative
signs of $B$, respectively.
Second, we have generalized these embeddings to the charged ST theories 
with the definitely positive $B$.
Third, we have also obtained the (3+3)-dimensional GEMS of the uncharged
and charged ST theories with the definitely positive parameter $L$.
Since in the uncharged limit $Q=0$, the (3+3)-dimensional 
coordinate transformations of the charged ST theories are exactly reduced 
to the uncharged case having the same (3+3)-dimensional GEMS structure 
in contrast to the usual BTZ case \cite{hkp}.  
Especially, since in the case with $\phi = r/(r-3B/2)$ the metric is 
Schwarzschild-like, the GEMS structure coinsides with that of the 
(3+1)-dimensional Schwarzschild black hole, which needs (1+1) additional 
extra dimensions to yield the (4+2)-dimensional GEMS structure.  
Furthermore, in the $B=0$ or $L=0$ limit, the coordinate transformations 
are exactly reduced to the charged BTZ case, 
which still has the (3+3)-dimensional GEMS structure.

\vskip 0.5cm

We acknowledge financial support in part from Ministry of Education, BK21
Project No. D-0055, 1999, and the Sogang University Research Grants in 1999.





\end{document}